\documentclass[%
 reprint,
 amsmath,amssymb,
 aps,
]{revtex4-2}

\usepackage{graphicx}
\usepackage{dcolumn}
\usepackage{bm}

\begin{document}

\preprint{APS/123-QED}

\title{Endoplasmic Reticulum Structure Determines Optimal Ribosome Density}

\author{Benjamin Tang}
\thanks{Contact author: bytang@stanford.edu}%
 \affiliation{Department of Chemistry, 290 Jane Stanford Way, S385, Stanford University, Stanford, CA 94305-4401, USA.}

\date{\today}

\begin{abstract}
Recent work has shown an increasing interest in understanding the structure of the endoplasmic reticulum (ER) and how ribosomes are displayed on it. Here we present a model that explains a physical reason for why the cell creates different structures of the ER. Due to the diffusion of biomolecules, we find that flat sheets and a matrix of tubules have different regimes of optimized capture efficiency. We extend the model to explain the observed difference in density of ribosomes on the structures of the ER. Due to the capture efficiency of tubules, less ribosomes are needed on those structures. For flat sheets, more ribosome coverage at biological separation distance is needed to match the same fraction of relative flux. We then push the model to predict that depending on the future life of the translated protein and overall demand for protein expression, the cell will utilize one structure of the ER over another. Predictions are compared with experimental data. 
\end{abstract}

\maketitle

There has been a growing interest in understanding the complex network of sheets and tubules that makes up the endoplasmic reticulum (ER) \cite{West2011}\cite{Scott2023}\cite{Puhka2012}. Specifically, why a cell would create different structures for ribosomes to anchor onto is of great interest. In fact, the structure of the rough ER has been shown to exist on a spectrum between planar sheets, rough ER, to a tubular matrix, smooth ER \cite{Puhka2012}. Moreover, it has been observed that the density of ribosomes on the ER depends heavily on the local curvature, with flat sheets having more ribosomes per area than tubules \cite{West2011}. In this letter, we present a model that provides a physical explanation for these observations. We then further push the model to predict which ER structure the cell will use depending on the eventual target of the protein and overall demand for protein expression. 

First, we will tackle the problem of differing structures to the ER. We begin with the geometry of the planar sheet. Inspired by Berg and Purcell (BP), we will treat the sheet as a rectangle with length $L$ and width $W$ in an infinite media, matching experimental observations \cite{Shibata2006}. For simplicity, assume $L\approx W$. 

Let a species $A$ have diffusion constant $D$ and bulk concentration $c_\infty$. Due to the complexity of finding the capacitance $C$ of such a geometry, especially later on, we take the limit where the sheet behaves similar to a disk of radius $R$ where $2R \approx L$. Then we can use Berg and Purcell's result, in CGS units, to find $C_{\mathrm{disk}}=2R/\pi$. This gives us that the flux of particles hitting the sheet is approximately $J_{\mathrm{sheet}}\approx J_{\mathrm{disk}} = 8RDc_\infty$. 

Next we introduce $N$ circular ribosomes with radius $r$ onto the sheet. We choose the model the ribosomes as circles on the sheet instead of spheres because $r<<R$. Let any molecule of $A$ that touches a binding site be immediately captured and then ejected, clearing the site for another catch. The surface of the sheet does not bind to species $A$. Assuming the ribosomes are somewhat uniformly distributed, we find a couple of notable results. First, the two limits are if there is one ribosome or the sheet is completely covered. These give $J_1 = 4Drc_\infty$ and $J_{\mathrm{max}} = J_{\mathrm{disk}}$. 

To tackle the intermediate case, we use the capacitance approach of BP for our geometry. For simplicity, assume that the radius of the ribosome disks is much smaller than the distance between the disks. This allows us to use nearest neighbor dominant assumption when finding the capacitance of the associated conductor surface. 

With these assumptions in mind, we find that our solution is similar to BP except that there is a term $\beta$ that comes out of the summation of Green's functions between disks because our geometry is not as symmetric as a sphere. Physically, this $\beta$ term acts as an intra-sheet interaction term in the high coverage limit. In fact, if we have $\beta = 1 $ as for a sphere, this result agrees with BP's finding as expected. 

Thus, we find the capacitance of such a covered sheet is
\begin{equation}
    C_{\mathrm{rib,disk}} =\frac{NrR}{\beta Nr + \pi R},
\end{equation}
where $\beta = \pi/2$ as found by fitting the saturation limit to the standard single disk equation. 

Now we can find the relative flux that these ribosomes take in
\begin{equation}
    \frac{C_{\mathrm{rib, disk}}}{C_{\mathrm{disk}}}=\frac{J_{\mathrm{rib, disk}}}{J_{\mathrm{disk}}} = \frac{Nr}{Nr + 2R}.
\end{equation}
Equipped with this equation, we find how the flux of absorption of the ribosomes quickly reaches a large portion of $J_{\mathrm{max}}$ without covering most of the sheet [Fig. 1(a), (b)]. In fact, when using biologically relevant values, such as $r=5$ nm and $R=1$ um, we find that at one percent coverage, the system reaches half saturation. 

Next, we move to modeling the double sheet geometry. We layer two parallel sheets with spacing $d$, which agrees with experimental observation of the rough ER \cite{Shibata2006}. The reason we can treat a stack of parallel sheets of the rough ER as just two is because most of the flux lines that penetrate one pair will penetrate the others, especially at small separation lengths. Thus, to first order, we choose to focus on two sheets in order to avoid the large but finite sum that would come with modeling the entire rough ER. If the sheets are much larger than the spacing $d<<R$, we use the standard result $C_{\mathrm{ppc}} = R^2/4d$. It is known that if the spacing is on the order of the size of the sheets, then there is no closed form analytical solution. Thus, we use the approximate solution \cite{Norgren2009}

\begin{equation}
    C_{\mathrm{ppc}} = \frac{1}{4\pi} \cdot [\frac{\pi R^2}{d} + R(\ln(\frac{R}{d} + \ln(16\pi)-1))]\approx \frac{R^2}{4d},
\end{equation}
keeping in mind that in the limit that $d\approx R$, we expect error on the order of $\ln(1/d)$. Fortunately, as we will see later, the biologically relevant regime of interest is $d<<R$. 

\begin{figure}[b]
\includegraphics[width=8.6cm]{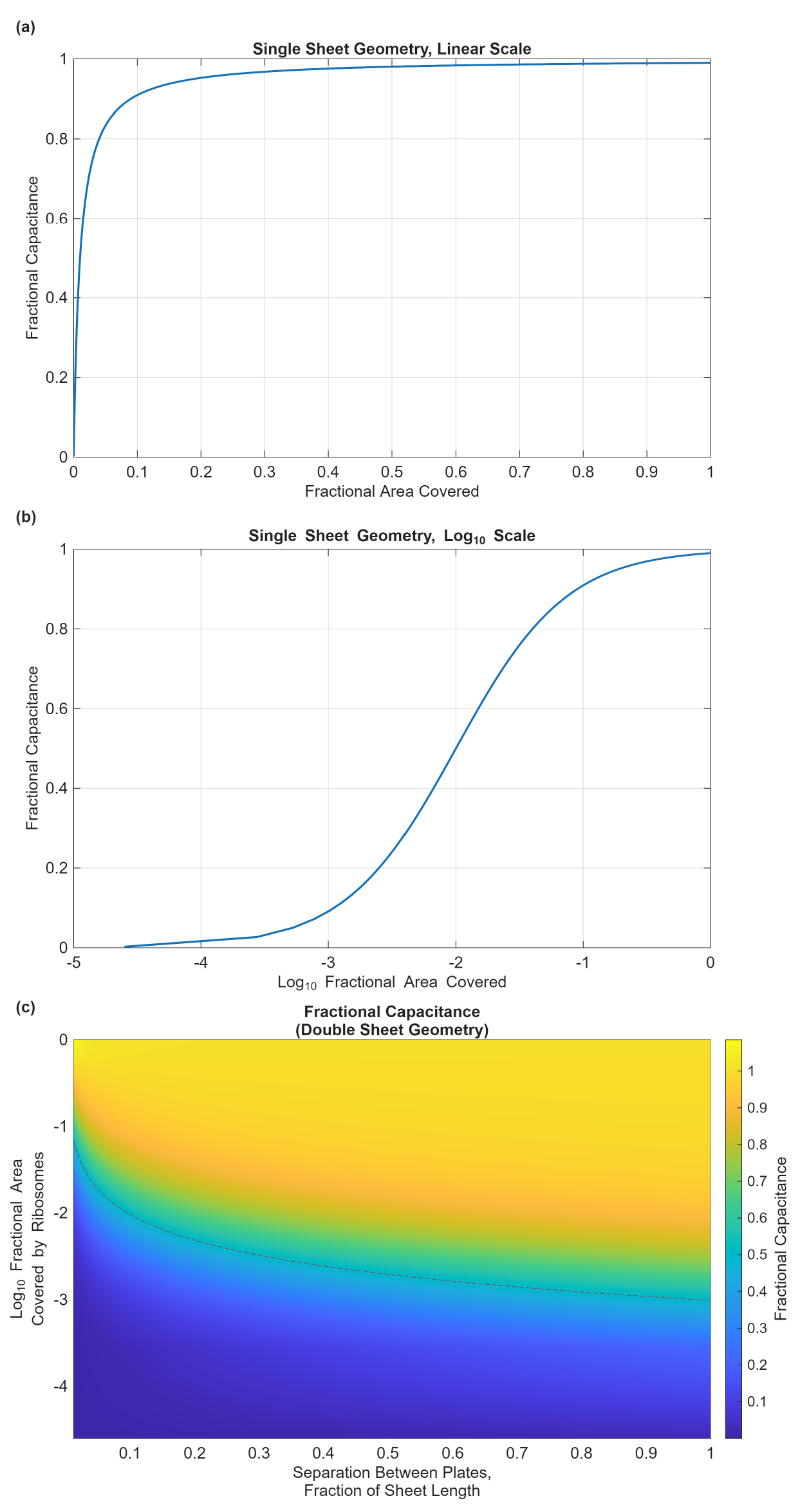}
\caption{\label{fig:epsart} (a) Plot showing the dependence of fractional capacitance for the single sheet geometry on fractional area covered by ribosomes. As expected, a small fraction of area covered leads to near saturation. (b) Same plot as in (a) except with $\log_{10}$ scaling. (c) Heatmap of the fractional capacitance for the double sheet geometry as a function of both separation and fractional area covered. Red contour shows where fractional capacitance is half saturated.}
\end{figure}

Next, we again introduce N ribosomes on each sheet with $r<<d$. Now there will be two interaction terms, one intra-sheet and one inter-sheet. We assume that the intra-sheet $\beta$ interaction term is the same as the single sheet. We label the inter-sheet interaction term $\gamma$. We introduce the ribosomes one-to-one on opposite planes, and then if the spacing between ribosomes is large as before, we have 
\begin{equation}
    C_{\mathrm{rib,sheets}} = \frac{NrR}{\beta Nr + R\pi + NrR\frac{\gamma}{d}}.
\end{equation}
Using the same $\beta$ interaction as before, we can again solve for $\gamma$ in the limit of saturation to get
\begin{equation}
    \gamma = \frac{d(8d-\pi (R+2r))}{2R^2}.
\end{equation}
Surprisingly, this inter-sheet interaction term scales with $d$. This occurs because $\gamma$ captures the screening one sheet has on the other, which for $d<<R$, grows linearly. Note we also expect this model to break down at extremely large $d$, as we would expect to recover the single sheet limit, which requires $\lim_{d\to\infty}\gamma(d)=0$. In that case, the parallel plate capacitor approximation we made would also break down.

We can then create the surface plot of the ratio of the ribosome covered sheets to the fully absorbing sheets [Fig. 1(c)]. We find that at extremely short distances where $r\approx d$, our model breaks down, as the ratio slightly exceeds 1. Thus, we choose to simulate at larger $d$ away from this limit. Beyond the extremely small separation limit, we reveal a telling result. We find that as distance between the sheets increases, the efficiency of the ribosome disks actually increases, meaning the system reaches saturation with less fractional area covered. 

Next, we use the ratio of $C_{\mathrm{rib,sheets}}$ to the full parallel plate capacitor as the relative flux captured by the ribosome covered sheets. We have 
\begin{equation}
    \frac{C_{\mathrm{rib,sheets}}}{C_{\mathrm{ppc}}} = \frac{4dNr}{R(\beta Nr + R\pi + NrR\frac{\gamma}{d})}.
\end{equation}

With this equation, we can find the difference in relative efficiencies of capture between the single and double sheet geometries (Fig. 2). We discuss two particularly interesting findings. First, we notice that at small separations, the single sheet saturates faster. This makes intuitive sense, as having two sheets in close proximity approaches a single sheet, yet it uses twice as many ribosomes. Additionally, the capacitance of the parallel plate capacitor that represents to total flux possible grows at smaller $d$. However, and most interesting, there is a distance and filling that allows the double sheet geometry to achieve a higher relative flux capture compared the single sheet. In particular, we find that as the separation $d>0.2L$ and $N \approx 0.01N_{\mathrm{max}}$ where $N_{\mathrm{max}}$ is the number of ribosomes to reach saturation, the double sheet transitions to having more efficiency than the single sheet. As the separation grows, the difference in efficiency continues to increase. 

These remarkable findings imply if the cell were to optimize only for relative efficiency, it would keep the rough ER sheets above a certain threshold distance apart, and it would pack an optimized amount of ribosomes per sheet in order to ensure efficient capture. However, it is experimentally known that the spacing between rough ER sheets is around $50$ nm \cite{Shibata2010}. Thus, in that regime, our model predicts in order for the double sheet to be as efficient as the single sheet, the cell would need approximately $4700$ ribosomes per $\mathrm{um}^2$, which is slightly larger than previous observations \cite{West2011}. Furthermore, note that the double sheet geometry has an higher absolute flux capture because $d<<R$ and
\begin{equation}
    \frac{C_{\mathrm{ppc}}}{C_{\mathrm{disk}}} = \frac{R\pi}{8d}.
\end{equation}

As for our prediction on the optimal spacing, we believe that while  $d>0.2L$ would provide greater relative advantage for the double sheet geometry, the cell uses the rough ER for other tasks, such as storing calcium. Thus, other biological factors require a denser packing than would be optimal based on our model.

\begin{figure}[b]
\includegraphics[width=8.6cm]{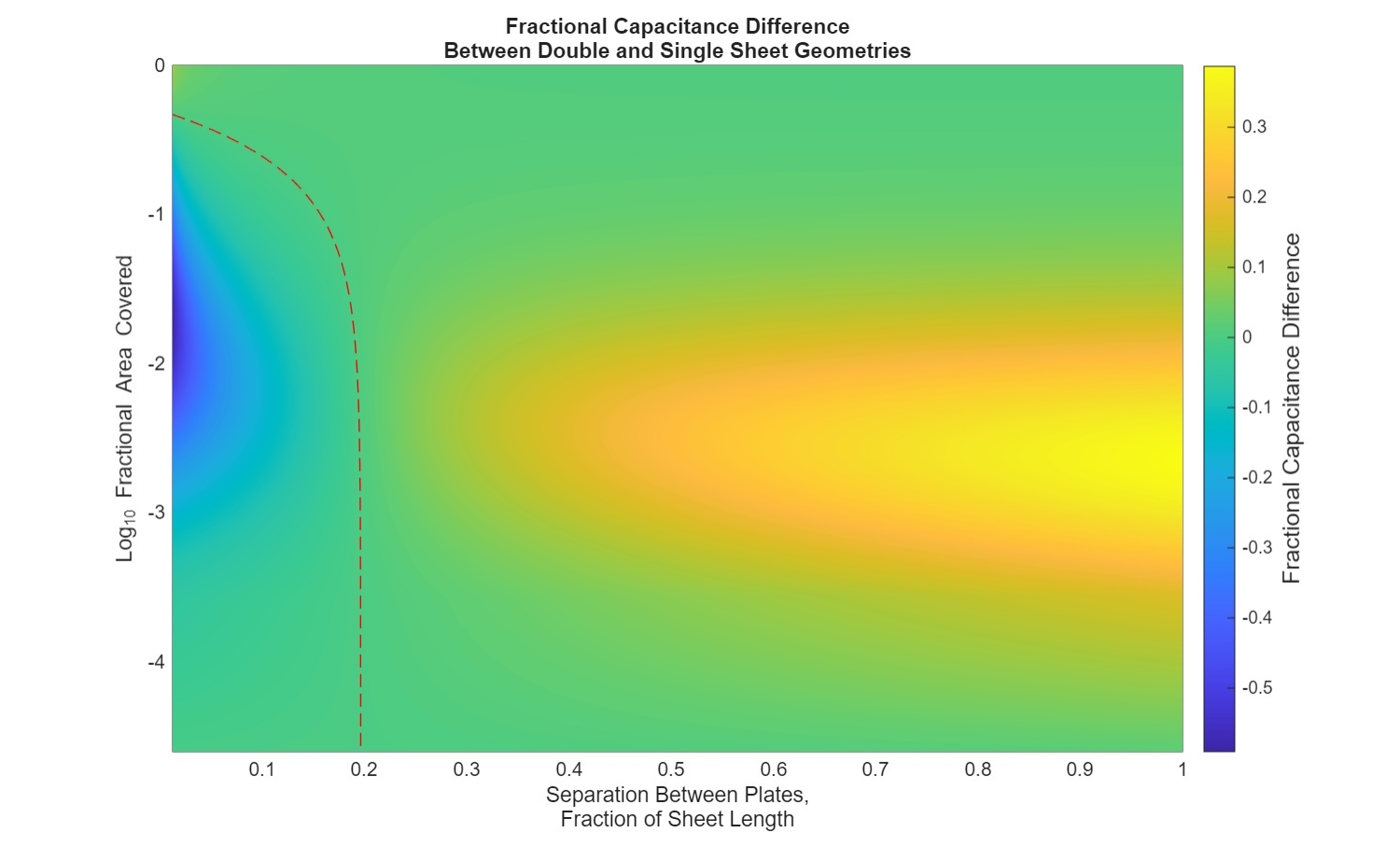}
\caption{\label{fig:epsart} Heatmap of the fractional capacitance difference between the double and single sheet geometries. There is a critical distance and filling such that the double sheet becomes more efficient. Red contour shows where fractional capacitance difference is zero.}
\end{figure}

We now move to modeling the tubular geometry. We can use the known capacitance of a tube with length $\ell$ and radius $a$ to get a capacitance of $C_{\mathrm{tube}} = \ell/2\ln(2\ell/a)$. Now we can compare the fluxes through the saturated disk and cylinder assuming the tube is made from wrapping up the sheet that the disk approximates. Because $a = \ell/2\pi$ and later we have $a<<\ell$, we ignore contributions from the faces of the cylinder. If we assume that $\ell\approx L\approx 2R$, then we have
\begin{equation}
    \frac{J_{\mathrm{disk}}}{J_{\mathrm{tube}}} = \frac{C_{disk}}{C_{tube}} = \frac{2R/\pi}{\frac{L}{2\ln(2L/a)}} = \frac{2\ln(4\pi)}{\pi} > 1
\end{equation}

This means that the sheet has a higher max flux than the cylinder, which intuitively makes sense. While tubules are much thinner in reality, this gives the same general result, that if the cell could have only one geometry for protein translation at full saturation, it would create sheets to maximize the probability of capture per unit area.

However, a cell is able to create multiple tubules below full saturation, so a more realistic model compares the relative fluxes between a stack of sheets and a matrix of tubules. Additionally, tubules have a circumference much smaller than the length of the sheets, meaning that for the same amount of surface area, the cell can create many more tubules. For our model, we treat each tube in the matrix as independent, as the separations between tubes is much larger than the separation between sheets biologically. 

First, we must find how the capacitance of a single tube saturates with the addition of $N$ ribosome disks, as with the other geometries. We derive an asymptotically controlled equation by matching the local disk solution to long-range cylindrical field results. This approach gives, for a tube of radius $a$ and length $\ell$, 
\begin{equation}
    C_{\mathrm{rib,tube}} = \frac{2\ell Nr}{\ell\pi + 4Nr\lambda \ln(\frac{2\ell}{a})}
\end{equation}

Again, we can find the inter-tube interaction term $\lambda$ for this geometry by solving in the fully saturated limit, and we find 
\begin{equation}
    \lambda = 1 - \frac{\pi r}{8a\ln(\frac{2\ell}{a})}
\end{equation}

Equipped with this, we can now find the relative flux captured by ribosomes on the tubular geometry
\begin{equation}
    \frac{C_{\mathrm{rib,tube}}}{C_{\mathrm{tube}}} = \frac{4Nr\ln(\frac{2\ell}{a})}{\ell\pi + 4Nr\lambda \ln(\frac{2\ell}{a})}
\end{equation}

Now we find the difference in fractional capacitance between tubular and double sheet geometries. For this simulation, we use known parameters for the tube, such as having a length $\ell=1$ um and a diameter of $2a=50$ nm \cite{West2011}. This gives rise to a heatmap that is similar to the one obtained from the double and single sheet geometry comparison, except with a few notable differences [Fig. 3(a)].

First, as the separation distance increases, the double sheet becomes more efficient. Thus, we find a critical distance below which the cell will opt to create a tubular matrix instead of a double sheet at a particular ribosome filling. However, again as noted before, the cell tends to avoid spacing out the sheets of the rough ER at the expected separation that maximizes its efficiency. Additionally, we keep in mind that the double sheet has higher absolute capacitance, and thus this is only an relative flux comparison. 

Now we use the model to explain the tighter packing of sheets observed. We see that the regime where the tubular geometry obtains higher efficiency than the double sheet is at a shorter separation distance than the single sheet comparison. In fact, the contour where the efficiencies between the double sheet and tubular geometries are equal remains at a separation of $100$ nm for a large range of fractional area covered.  

Next, we compare the relative capacitances between these two geometries using known experimental parameters [Fig. 3(b)]. This highlights the difference in relative efficiencies at biological relevant conditions. In fact, we then took the fractional area coverage that allows the relative efficiencies of 50 percent for both the tube and double sheet at a separation of $50$ nm. We find a ribosomal coverage density of $733$ ribosomes per $\mathrm{um}^2$ on the double sheet and a ribosomal coverage density on the tube is $272$ ribosomes per $\mathrm{um}^2$. Assuming the cell will try to maintain this efficiency on all of its ER surfaces, these findings fall within or near the bounds found experimentally \cite{West2011}\cite{Puhka2012}. 

These findings underscores the existence of another constraint that the cell faces when designing ER structures. The rough ER could achieve the same efficiency as the smooth ER at a separation of $100$ nm, yet it experimentally stacks more tightly. 

Recent work has shown that regions of high curvature offer a geometric reason for a higher density of flux lines compared to flat regions \cite{Wu2025}. We show that the same is true for the geometries presented here. 

Next, we consider how these different regions may impact the protein after translation. Considering the mean free path $x_{\mathrm{rms}}$ of particles diffusing in the double sheet geometry, we find to first order $x_{\mathrm{rms, sheet}} \approx d$. However, for the tubular matrix, we have $x_{\mathrm{rms, matrix}}\approx L$. Thus, for large number of tubules, the matrix offers less collisions per unit time than the double sheet. 

Since the tubular matrix offers longer mean free paths, this also implies that the protein takes less time to diffuse away from the matrix. On the other hand, the flat sheets offer a relatively slower escape for the proteins after translations. Of course, the cell utilizes bimolecular machines to overcome diffusion limited processes, but that requires energy. Thus, we hypothesize that the cell will translate rare proteins that require fast transport in the matrix and use active transport, and it will translate more common proteins on the rough ER, supported by the differences in absolute flux capture between the two geometries. Another possible solution to the diffusion problem that one study found is the formation of "hotspots" in the tubular matrix. These allow for faster pathways for particle diffusion, so we expect those secondary effects to play a role in a cell \cite{Scott2023}. We do not model this inhomogeneity. 

In conclusion, we have demonstrated, for the first time to our knowledge, that the geometry of sheets and tubules gives a physical explanation for the differences in ribosome densities. The rough ER, sheets, requires more ribosomes per area to saturate but offers a higher absolute flux capture and a slower exit path for translated proteins. The smooth ER, tubular matrix, allows for efficient use of ribosomes and faster diffusion times but has lower absolute flux capture. Furthermore, only a small fraction of the ER surface area needs to have ribosomes to achieve relatively high rates of flux capture. Of course, previous work also finds conflicting evidence that local curvature isn't the only factor determining ribosome density \cite{West2011}, which we may be related to the slight difference in predicted and measured separation distance between sheets in the rough ER. 

\begin{figure}[b]
\includegraphics[width=8.6cm]{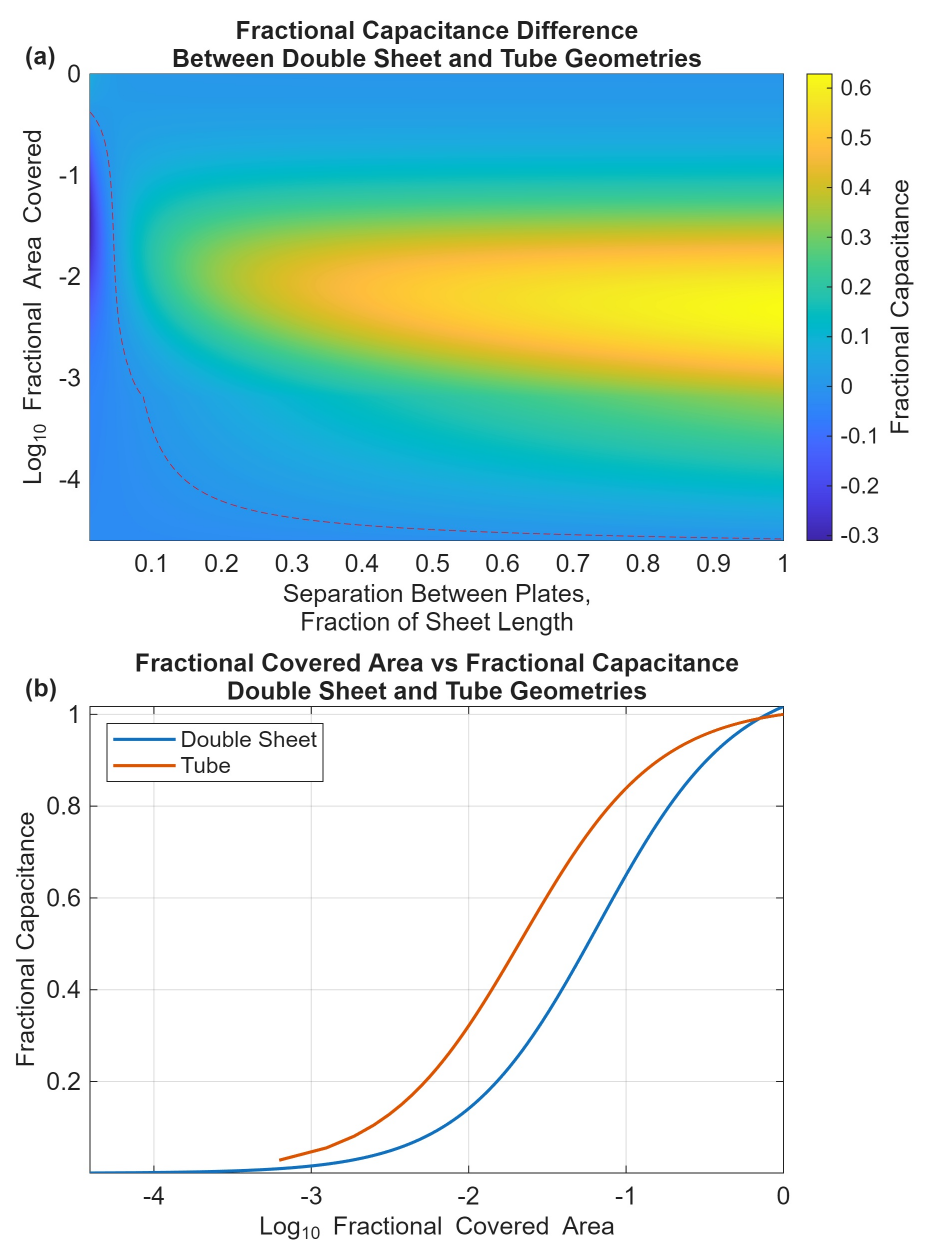}
\caption{\label{fig:epsart} (a) Heatmap of the fractional capacitance difference for a double sheet and fixed tube. Red contour shows where fractional capacitance difference is zero. (b) Fractional capacitance comparison between the double sheet and tubular geometries with known parameters.}
\end{figure}

\bibliographystyle{apsrev4-2}   
\bibliography{refs}             

\begin{acknowledgments}

This work was supported through the Chemistry Department at Stanford University. 
\end{acknowledgments}

\nocite{*}


\end{document}